\documentclass[aps,prl,twocolumn,superscriptaddress,nobibnotes,letterpaper]{revtex4}
\usepackage{times}
\usepackage[pdftex]{graphicx}
\usepackage[pdftex]{epsfig}
\usepackage{amsmath}
\usepackage{pslatex}
\usepackage{amssymb}
\usepackage{amsfonts}
\usepackage{color}
\usepackage{wrapfig}
\usepackage{eucal}
\usepackage{hhline}
\usepackage{threeparttable}
\usepackage{supertabular}
\usepackage{multirow}
\usepackage{tabularx}
\usepackage{setspace}
\usepackage[compact]{titlesec}
\usepackage{hyperref}
\usepackage{graphicx}
\usepackage{esint}
\usepackage{verbatim}
\usepackage{units}

\newcommand{\kkindex}{n}
\newcommand{\betaTH}{\beta_{\text{th}}}
\newcommand{\betam}{\beta_{m}}
\newcommand{\betaEOM}{\beta_{_\text{{EOM}}}}
\newcommand{\Tbath}{T_{b}}

\newcommand{\Hint}{H_{\text{int}}}

\newcommand{\smallprox}{ \! \approx \!}

\newcommand{\cto}{c_\text{to}}
\newcommand{\Deltal}{\Delta_{L}}
\newcommand{\Deltap}{\Delta_{p}} 
\newcommand{\omegap}{\omega_{p}}
\newcommand{\DeltalO}{\Delta_{L,0}}
\newcommand{\Deltam}{\Delta_{m}}
\newcommand{\ain}{a_{\text{in}}}
\newcommand{\Pin}{P_{\text{in}}}
\newcommand{\Pfric}{P_{\text{fric}}}
\newcommand{\Prad}{P_{\text{rad}}}
\newcommand{\lambdacO}{\lambda_{c}}

\newcommand{\ncavObar}{\bar{n}_{a}}
\newcommand{\gzeroO}{g_{0}}

\newcommand{\kappai}{\kappa_{i}}
\newcommand{\kappae}{\kappa_{e}}
\newcommand{\gammamO}{\gamma_{i}}  
  
\newcommand{\omegac}{\omega_{c}}

\newcommand{\omegamO}{\omega_{m}}
\newcommand{\omegam}{\omega_{m}}
\newcommand{\omegal}{\omega_{L}}

\newcommand{\xzpf}{x_{\text{zpf}}}

\newcommand{\mO}{\text{m}_{\text{eff}}}

\newcommand{\Figref}[1]{Fig.~\ref{#1}}
\newcommand{\Eqref}[1]{Eq.~\ref{#1}}

\newcommand{\Mag}[1]{|{#1}|^2}
\newcommand{\mutwo}{\mu}

\begin{document}

\pagenumbering{arabic}

\title{Nonlinear radiation pressure dynamics in an optomechanical crystal 
}

\author{Alex G. Krause}
\affiliation{Kavli Nanoscience Institute and Thomas J. Watson, Sr., Laboratory of Applied Physics, California Institute of Technology, Pasadena, CA 91125}
\affiliation{Institute for Quantum Information and Matter, California Institute of Technology, Pasadena, CA 91125}
\author{Jeff T. Hill}
\affiliation{Kavli Nanoscience Institute and Thomas J. Watson, Sr., Laboratory of Applied Physics, California Institute of Technology, Pasadena, CA 91125}
\affiliation{Institute for Quantum Information and Matter, California Institute of Technology, Pasadena, CA 91125}
\affiliation{Edward L. Ginzton Laboratory, Stanford University, Stanford, CA 94305}

\author{Max Ludwig}
\affiliation{Institute for Theoretical Physics, Universit\"at Erlangen-N\"urnberg, 91058 Erlangen}

\author{Amir H. Safavi-Naeini}
\affiliation{Kavli Nanoscience Institute and Thomas J. Watson, Sr., Laboratory of Applied Physics, California Institute of Technology, Pasadena, CA 91125}
\affiliation{Institute for Quantum Information and Matter, California Institute of Technology, Pasadena, CA 91125}
\affiliation{Edward L. Ginzton Laboratory, Stanford University, Stanford, CA 94305}

\author{Jasper Chan}
\affiliation{Kavli Nanoscience Institute and Thomas J. Watson, Sr., Laboratory of Applied Physics, California Institute of Technology, Pasadena, CA 91125}
\affiliation{Institute for Quantum Information and Matter, California Institute of Technology, Pasadena, CA 91125}

\author{Florian Marquardt}
\affiliation{Institute for Theoretical Physics, Universit\"at Erlangen-N\"urnberg, 91058 Erlangen}
\affiliation{Max Planck Institute for the Science of Light, G\"unther-Scharowsky-Stra\ss e 1/Bau 24, D-91058 Erlangen, Germany}

\author{Oskar Painter}
\email{opainter@caltech.edu}
\homepage{http://copilot.caltech.edu}
\affiliation{Kavli Nanoscience Institute and Thomas J. Watson, Sr., Laboratory of Applied Physics, California Institute of Technology, Pasadena, CA 91125}
\affiliation{Institute for Quantum Information and Matter, California Institute of Technology, Pasadena, CA 91125}

\date{\today}

\begin{abstract}
Utilizing a silicon nanobeam optomechanical crystal, we investigate the attractor diagram arising from the radiation pressure interaction between a localized optical cavity at $\lambdacO = 1552$~nm and a mechanical resonance at $\omegamO/2\pi = 3.72$~GHz. At a temperature of $\Tbath \approx 10$~K, highly nonlinear driving of mechanical motion is observed via continuous wave optical pumping.  Introduction of a time-dependent (modulated) optical pump is used to steer the system towards an otherwise inaccessible dynamically stable attractor in which mechanical self-oscillation occurs for an optical pump red-detuned from the cavity resonance.  An analytical model incorporating thermo-optic effects due to optical absorption heating is developed, and found to accurately predict the measured device behavior.
\end{abstract}


\maketitle

The field of optomechanics, concerned with the interaction of an optical cavity and a mechanical resonator~\cite{kippenberg_cavity_2008}, has been of recent interest for its promise for use in sensors \cite{krause_high-resolution_2012,srinivasan_optomechanical_2011}, nonlinear optics~\cite{hill_coherent_2012,safavi-naeini_squeezed_2013}, and demonstrations of macroscopic quantum mechanics~\cite{chen_macroscopic_2013,marshall_towards_2003}. To lowest order, the mechanical displacement linearly modulates the frequency of the optical resonance in a cavity-optomechanical system.  This, however, gives rise to an inherently nonlinear phase modulation, and through radiation-pressure backaction on the mechanical element, yields nonlinear system dynamics~\cite{marquardt_dynamical_2006}. Much of the previous work has focused on the linearized regime where the interaction with the optical field still gives rise to a host of interesting phenomena such as a modified spring constant \cite{eichenfield_picogram-_2009}, damping or amplification of the mechanics \cite{lin_mechanical_2009}, and EIT-like slow-light effects \cite{safavi-naeini_electromagnetically_2011,weis_optomechanically_2010}. Recently, several experiments have pushed into the quantum regime using back-action cooling in the linearized regime to damp nanomechanical resonators to near their quantum ground state of motion~\cite{chan_laser_2011,teufel_sideband_2011}.

\begin{figure}
 \includegraphics[scale=1, clip= true]{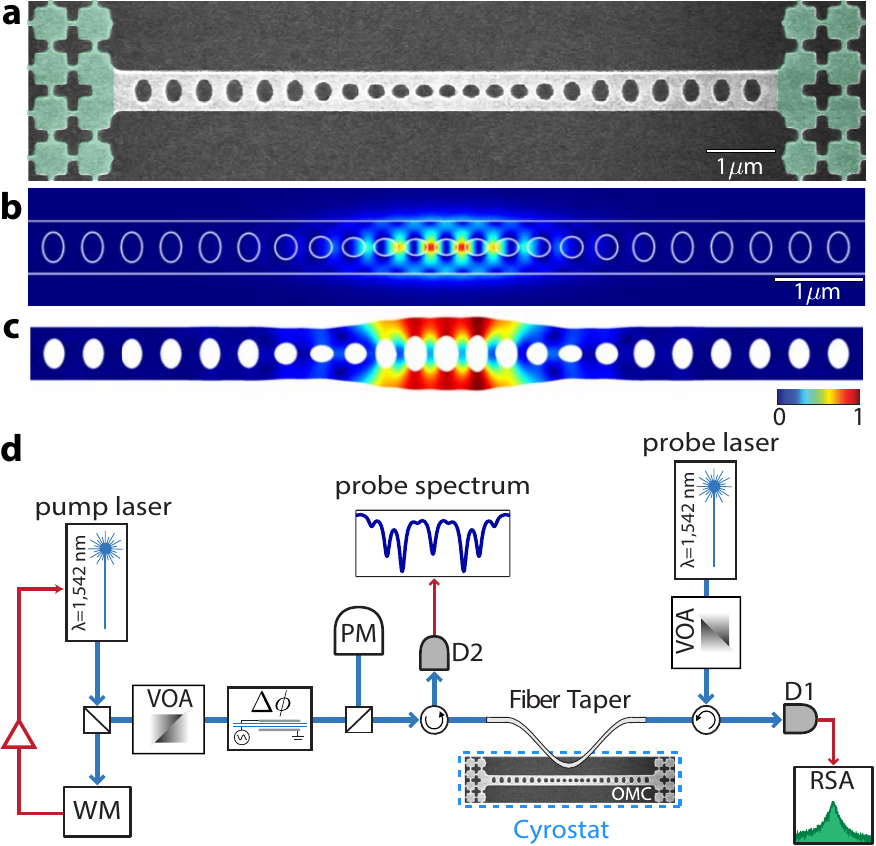}
 \caption[]%
  {(a) SEM of the optomechanical nanobeam surrounded by phononic shield (green). (b) FEM-simulated electromagnetic energy density of first-order optical mode; white outline denotes edges of photonic crystal. (c) FEM-simulated mechanical mode profile (displacement exaggerated). In (b) the colorscale bar indicates large (red) and small (blue) energy density, whereas in (c) the scale bar indicates large (red) and small (blue) displacement amplitude. (d) Simplified schematic of experimental setup. WM: wavemeter, $\Delta\phi$: electro-optic phase modulator, OMC: optomechanical crystal, D1: pump light detector, D2: probe detector, VOA: variable optical attenuator, PM: power meter}\label{device}
\end{figure}

In this work, we instead demonstrate new features and tools in the nonlinear regime of large mechanical oscillation amplitude. Previous experimental works have shown that a blue-detuned laser drive can lead to stable mechanical self-oscillations~\cite{carmon_temporal_2005,rokhsari_radiation-pressure-driven_2005,kippenberg_analysis_2005,karrai_doppler_2008}, or even chaotic motion~\cite{carmon_chaotic_2007}. Theoretical predictions of an intricate multistable attractor diagram~\cite{marquardt_dynamical_2006} have so far eluded experimental observation, except for the elementary demonstration of dynamical bistability in a \text{photothermally} driven system~\cite{metzger_self-induced_2008}. In the present work, we are able to verify the predicted attractor diagram and further utilize a modulated laser drive to steer the system into an isolated high-amplitude attractor. This introduces pulsed control of nonlinear dynamics in optomechanical systems dominated by radiation pressure backaction, in analogy to what has been shown recently for a system with an intrinsic mechanical bistability~\cite{bagheri_dynamic_2011}.


\begin{figure*}
 \includegraphics[width= 2\columnwidth]{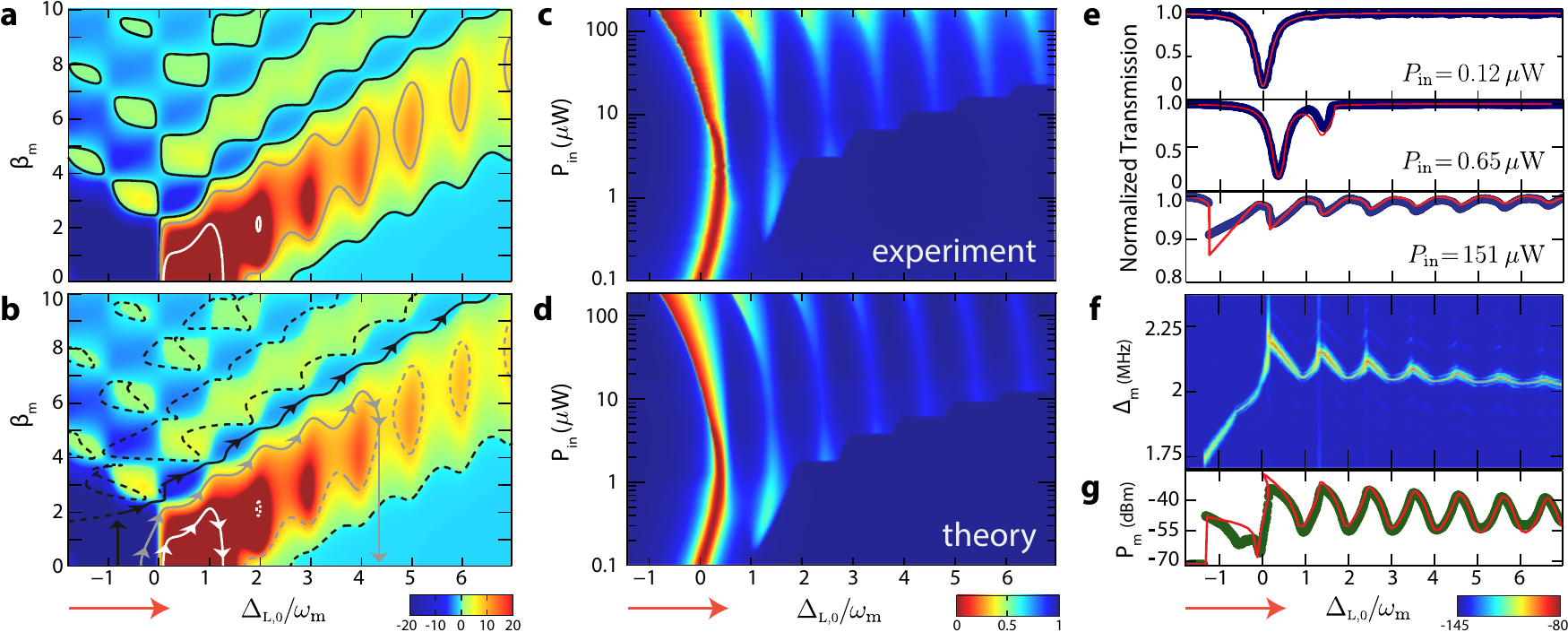}
 \caption[Gain Spectrum and Thermal Bistability Shift]%
  {(a) Calculated gain spectrum for the OMC in the amplitude-detuning plane. Color scale indicates the ratio of power input to that lost from friction $\left(\Prad/\Pfric - 1\right)$ at $\Pin = 151$~$\mutwo$W. Positive values are regions of mechanical self-oscillation. Solid line curves indicate power-conserving solution contours at selected input powers:  $0.65$~$\mutwo$W (white), $6.5$~$\mutwo$W (grey), $151$~$\mutwo$W (black). (b) Same as (a) with contours now shifted by estimated thermo-optic effects.  Solid line curves indicate the path taken by the mechanical oscillator during the laser sweep. Dashed lines are contours which are either unstable or unreachable by a slow sweep of laser detuning. (c) Image plot of the measured optical transmission spectrum versus laser detuning and power. (d) Image plot of the theoretically calculated transmission spectra including thermo-optic shifts and a slow drift in the optical resonance frequency over the course of the measurement from low to high power.  Spectra in (c) and (d) are scaled at each power level to span the range $0-1$. (e) Plot of the normalized optical transmission from scans in (c) at $\Pin = 0.12\ \mutwo$W (top), $0.65\ \mutwo$W (center), $151\ \mutwo$W (bottom).  Blue points are measured data and red curve is the theoretical model. (f) Power spectral density of detected signal near the mechanical frequency for $\Pin = 151\ \mutwo$W, showing frequency shifts of the mechanical mode, $\Deltam \equiv \omegamO - \omega_{m,0}$, from its bare frequency $\omega_{m,0}/2\pi = 3.72$~GHz. Color scale is detected power density in dBm/Hz. (g) Total integrated power of spectra in (f). Measured data are plotted as green circles, with the theoretical model (up to a scale factor) shown as a solid red curve.  The red arrow in each plot indicates the laser scan direction.
}\label{waterfall}
\end{figure*}

We employ a one-dimensional (1D) optomechanical crystal (OMC) designed to have strongly interacting optical and mechanical resonances~\cite{chan_optimized_2012}. The OMC structure is created from a free-standing silicon beam by etching into it a periodic array of holes which act as Bragg mirrors for both acoustic and optical waves~\cite{eichenfield_optomechanical_2009}. A scanning electron micrograph (SEM) of an OMC cavity is shown in \Figref{device}a along with finite-element-method (FEM) simulations of the co-localized optical~(\Figref{device}b) and mechanical~(\Figref{device}c) resonances. To reduce radiation of the mechanical energy into the bulk, the OMC is surrounded by a periodic `cross' structure which has a full acoustic bandgap around the mechanical frequency (\Figref{device}a, green overlay)~\cite{alegre_quasi-two-dimensional_2011}.

The experimental setup is shown schematically in \Figref{device}d. The silicon chip containing the device is placed into a helium flow cryostat where it rests on a cold finger at $T\approx4$~K (the device temperature is measured to be $\Tbath \approx 10$~K). Input laser light is sent into the device via a tapered optical fiber, which, when placed in the near-field of the device, evanescently couples to the optical resonance of the OMC~\cite{michael_optical_2007}. The transmitted light is detected on a high-bandwidth photodiode (D1) connected to a real-time spectrum analyzer (RSA). We also employ an electro-optic modulator (EOM) in the laser's path to resonantly drive the mechanical resonator. Finally, we can send in a low-power, counter-propagating probe laser whose detected spectrum (D2) is used to measure the mechanical amplitude and the pump-cavity detuning.  Using this set-up the optical resonance of the device studied in this work is measured to be at at $\lambdacO = 1542$~nm, with intrinsic (taper loaded) energy decay rate of $\kappai/2\pi= 580$~MHz ($\kappa/2\pi=1.7$~GHz). The breathing mechanical mode, shown in \Figref{device}c, is found to be at $\omegamO/2\pi = 3.72$~GHz with bare energy damped rate of $\gammamO/2\pi = 24$~kHz.  


The coupling of the optical resonance frequency to the mechanical displacement yields the interaction Hamiltonian, ${\Hint = \hbar \gzeroO\hat{a}^{\dagger}\!\hat{a}\hat{x}}$ where $\hat{a}$ ($\hat{x}$) is the optical (mechanical) field amplitude, and $\gzeroO$ is the vacuum optomechanical coupling rate. The physical mechanical displacement expectation is given by $x=\xzpf \langle \hat{x} \rangle$, where the zero-point amplitude of the resonator is $\xzpf=\left(\hbar/2\mO\omegamO\right)^{1/2} = 2.7$~fm (estimated using a motional mass $\mO = 311$~fg calculated from FEM simulation). Utilizing a calibration of the per-photon cooling power~\cite{chan_laser_2011} we find that $\gzeroO/2\pi = 941$~kHz. These device parameters put our system well into the sideband resolved regime $\kappa/\omegamO \ll 1$.



The classical nonlinear equations of motion for the mechanical displacement ($x$) and the optical cavity amplitude ($a = \langle{\hat a}\rangle$) are,

\begin{align}
	\ddot{x}(t)=-\gammamO \dot{x}(t) - \omegamO^2 x\left(t\right)+2\omegamO \gzeroO\xzpf \Mag{a(t)}, \\
	\dot{a}(t)=\left[ -\frac{\kappa}{2} +i\left(\Deltal+ \frac{\gzeroO}{\xzpf} x(t)\right) \right]a(t) + \sqrt{\frac{\kappae}{2}} \ain,
\end{align}

\noindent where $\ain = \sqrt{\Pin/\hbar\omegal}$ is the effective drive amplitude of the pump laser (input power $\Pin$ and frequency $\omegal$), $\kappae/2$ is the fiber taper input coupling rate, $\omegac$ is the optical cavity resonance frequency, and $\Deltal \equiv \omegal-\omegac$.  Since we are interested in the regime of self-sustained oscillations, where the motion of the oscillator is coherent on time scales much longer than the cavity lifetime, we can take the mechanical motion to be sinusoidal with amplitude $A$, ${x(t) = A\sin{\omegamO t}}$. The optical cavity field is then given by,

\begin{align}
   a\left(t\right) =  \sqrt{\frac{\kappae}{2}} \ain e^{i\Phi\left(t\right)} \sum_\kkindex{i^{\kkindex} \alpha_{\kkindex} e^{i\kkindex\omegamO t} }, \label{afinal}
\end{align}

\noindent where $\Phi\left(t\right) = -\betam \cos{\omegamO t}$ and $\alpha_{\kkindex} = J_\kkindex \left(\betam \right)/\left(\kappa/2 + i\left(\kkindex\omegamO-\Deltal\right)\right)$.  Here $J_\kkindex$ is the Bessel function of the first kind, $\kkindex$-th order, and its argument is the unitless modulation strength $\betam = \left(A\gzeroO\right)/\left(\xzpf\omegamO\right)$.  For $\betam \ll 1$ only the terms oscillating at the mechanical frequency, $\omegamO$, are appreciable, so the interaction can be linearized, and only the first-order radiation pressure terms are present. However, for $\beta \ge 1$ the higher harmonic terms at each~$\kkindex\omegamO$ have significant amplitude and backaction force.

The thermal amplitude is too small to enter the nonlinear regime in our devices $(\betaTH \approx 0.01)$, however, backaction from the pump laser can provide amplification to drive the mechanical resonator into the high-$\beta$, nonlinear regime. The resulting mechanical gain spectrum in the amplitude-detuning plane (the attractor diagram) can be solved for by calculating the energy lost in one mechanical cycle ($\Pfric = \mO \gammamO \left<\dot{x}^2\right>$) and comparing it to that gained (or lost) from the optical radiation force ($\Prad = \left(\hbar\gzeroO/\xzpf\right)\left<\Mag{\hat{a}}\dot{x}\right>$)~\cite{marquardt_dynamical_2006} .  Figure~\ref{waterfall}a shows a plot of the gain spectrum for the parameters of the device studied here with a laser pump power of $\Pin=$151~$\mutwo$W. Imposing energy conservation, $\Prad/\Pfric = +1$, yields the steady-state solution contour lines. Although the entire contour is a physical solution, the equilibrium is only stable when the power ratio decreases upon increasing the mechanical amplitude, $\frac{\partial}{\partial \beta} \frac{\Prad}{\Pfric}< 0$ (i.e. stability is found at the 'tops' of the contours)~\cite{marquardt_dynamical_2006}.

In the device studied here there is a thermo-optic frequency shift of the optical cavity caused by heating due to intra-cavity optical absorption. The thermal time constant of the device structure is slow relative to the optical cavity coupling rate, but fast compared to the laser scan speed.  Absorption heating can thus be modeled as a shift of the laser detuning proportional to the average intra-cavity photon number ($\ncavObar$), $\Deltal = \DeltalO + \cto\ncavObar$, where $\DeltalO$ is the bare laser-cavity detuning in absence of thermo-optic effects.  The per photon thermo-optic frequency shift of the optical cavity is measured to be $\cto/2\pi = -216$~kHz. Including this effect, the shifted contours are shown in \Figref{waterfall}b as a function of the bare detuning $\DeltalO$. The solid lines with arrows indicate the expected path traversed by the mechanical resonator during a slow laser scan from lower to higher laser frequency (left to right) at each power. The dashed lines are contours that are either unstable, or unreachable by this adiabatic laser sweep.

We first explore the lowest-lying contour of the attractor diagram by measuring the optical transmission as the pump laser is tuned from red to blue across the optical cavity resonance with a fixed optical input power.  A dip in transmission indicates that light is entering the cavity and being lost through absorption or scattering. At low optical input powers (${\Pin < 0.3\ \mutwo}$W), only a single resonance dip associated with the bare optical cavity is observed. Upon increasing the laser power, radiation pressure backaction amplifies the thermal motion of the mechanical resonator beyond threshold and into a large amplitude state. When this occurs a large fraction of the intra-cavity photons are scattered, resulting in additional transmission dips near each detuning $\DeltalO = \kkindex\omegam$ where threshold is reached.  Physically, mechanical oscillations at the $\kkindex$-th sideband detuning are generated by a multi-photon gain process involving $\kkindex$ photon-phonon scattering events. This stair-step behavior is seen in the measured transmission spectrum of both \Figref{waterfall}c and \Figref{waterfall}e. The theoretically calculated spectra including thermo-optic effects are shown in \Figref{waterfall}d, in excellent agreement with the measured spectra after taking into account a slow drift in the cavity resonance frequency as the measurements were taken from low to high power.

\begin{figure*}
 \includegraphics[width = 2\columnwidth]{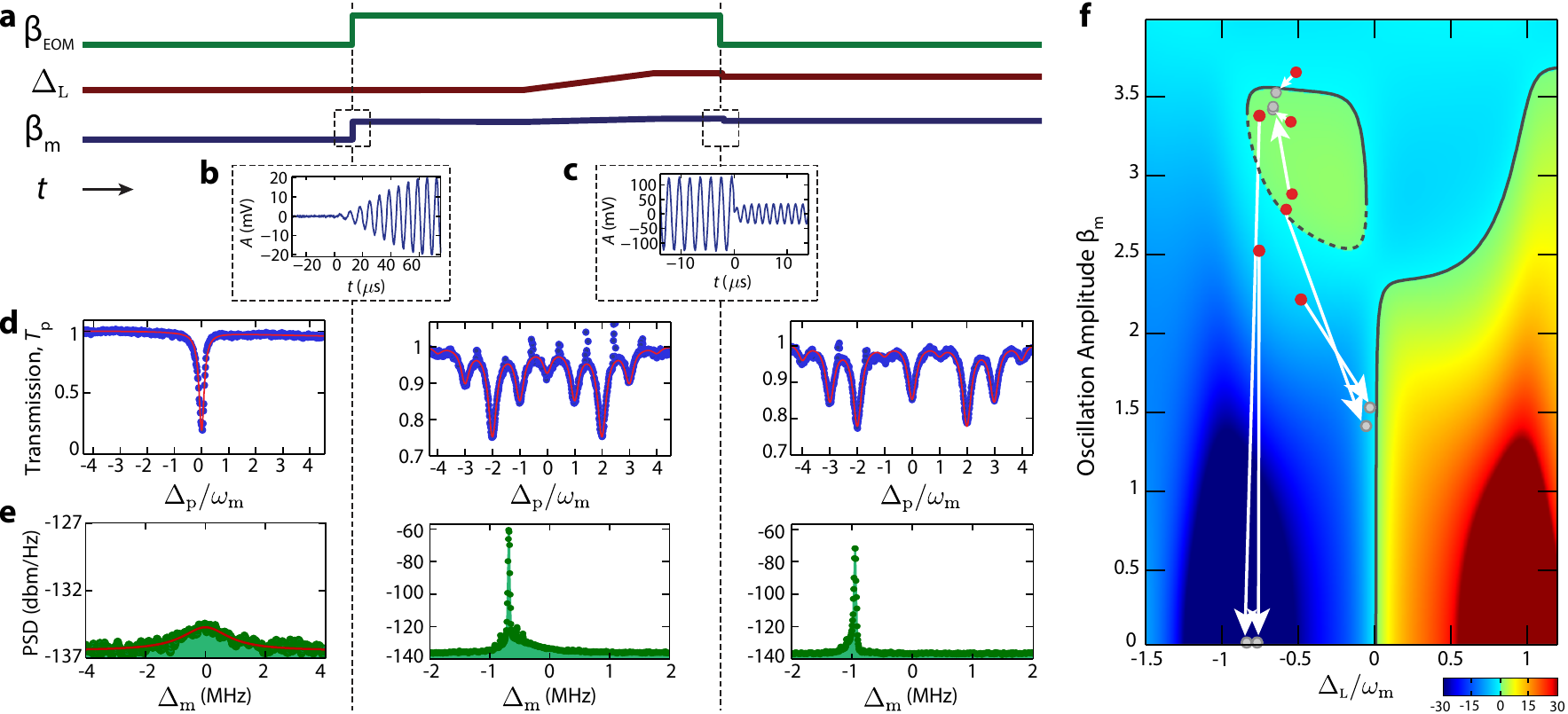}
 \caption[Transmission and Sideband Amplitude Fits]%
  {(a) Schematic showing the state preparation and measurement sequence used to explore the higher lying attractors. (b) Measured time-domain signal of one quadrature of the mechanical amplitude during the turn-on of the EOM drive ($t=0$). For clarity, the signal is mixed down from the mechanical resonance frequency at $3.7$~GHz to $150$~kHz. (c) Measured mechanical amplitude (mixed down to $500$~kHz) during the turn-off of EOM drive modulation at $t=0$.  Here we show a case where the system remained trapped in the high-amplitude state. (d) Transmission scans of the counter-propagating weak probe laser during each sequence of the measurement.  Blue points are measured data and red curves are fits to the data using \Eqref{afinal}. $\Deltap \equiv \omegap - \omegac$. (e) Mechanical power spectrum (green points) near the mechanical frequency during each sequence of the measurement. In the left panel the resonator is in a cooled thermal state and the red curve is a fit used to extract mechanical resonator parameters. (f) Plot of the normalized gain spectrum in the detuning-amplitude plane with overlaid stable solution contour (black solid curve) at $\Pin = 43~\mutwo$W. Color scale is $\left(\Prad/\Pfric - 1\right)$. Dashed black curve indicates unstable portion of contour. Red data points indicate initial $(\betam,\Deltal)$ and grey data points indicate the final values, for selected instances of the experimental sequence as shown in (a). White arrows connect initial/final pairs but do not indicate the actual path taken by the system.} \label{RSL}
\end{figure*}

Figure~\ref{waterfall}f shows the radio-frequency noise power spectrum near the mechanical frequency of the optical transmission photocurrent at the highest measured input power ($\Pin = 151\ \mutwo$W).  We note that backaction effects blue-shift the mechanical resonance frequency by an appreciable amount ($\sim 2$~MHz) from its intrinsic value of $\omegam/2\pi=3.72$~GHz. This frequency shift, $\Deltam$, depends on the laser detuning in a more intricate fashion than a linear calculation of the optical spring effect would suggest. A measure of oscillation amplitude can be extracted from the total power in this mechanical sideband, and its measured dependence on detuning is shown in \Figref{waterfall}g.  The measured total transduced power oscillates due to the nonlinearity of the detection process, which is nicely captured by the theoretical model, obtained without fit parameters except for an overall scale factor (\Figref{waterfall}g, red curve).



It is readily apparent from \Figref{waterfall}a that at large optical powers (black contour) there are a number of isolated attractor contours at higher oscillation amplitudes. Here we utilize external time-dependent driving of the mechanical mode to explore the lowest-lying isolated attractor on the red side of the optical cavity ($\Deltal < 0$), where the linearized theory predicts only damping of the mechanical mode.  An electro-optic modulator (EOM) is utilized to phase-modulate the incoming light field, resulting in an oscillating force inside the cavity which drives the mechanical resonator towards higher amplitudes.  The experimental sequence is displayed in \Figref{RSL}a.  We start with the pump laser switched on at a power of ${\Pin = 43\ \mutwo}$W, the laser detuned to the red side of the cavity resonance, and the phase modulation off ($\betaEOM = 0$), which initializes the mechanical resonator into a cooled thermal state with $\betam\approx 0$. The EOM phase modulation is then turned on which rings up the mechanical resonator.  Following ring-up, the pump laser is tuned to a starting detuning of $\Deltal$, completing the initialization sequence. Finally, the modulation is switched off ($\betaEOM \rightarrow 0$) and the system is allowed to relax into a final mechanical oscillation amplitude and laser-cavity detuning.  

In order to determine $\betam$ and $\Deltal$ at each stage of the above procedure an additional counter-propagating weak optical probe beam of frequency $\omegap$ is scanned across the cavity (see panels in \Figref{RSL}d). When the mechanical amplitude is large ($\betam \gtrsim 1$) the standard single resonance dip is transformed into a multi-featured spectrum, with resonance dips at the motional sidebands ($\Deltap \equiv \omegap-\omegac = \pm \kkindex \omegam$). A fit to the probe spectrum is performed using \Eqref{afinal}, with $\betam$ and $\omegac$ as free parameters (\Figref{RSL}d, red curves).  Repeating the measurement for different initial states and recording the resulting final states reveals the flow in the underlying attractor diagram.  A representative subset of the results are plotted in \Figref{RSL}f. We find that for a narrow range of initial conditions, after the modulation is switched off the system remains trapped close to the predicted top of the higher amplitude attractor at $\betam \smallprox 3.5$ and $\Deltal/\omegamO \smallprox -0.7$. For more negative initial detunings or lower initial mechanical amplitudes, the system relaxes into the trivial low-amplitude state or gets caught on the lowest-lying contour explored in \Figref{waterfall} (for detunings $\Deltal/\omegamO > -0.5$ the system could not be stably initialized due to the thermo-optic effect).  





The results presented here represent an initial exploration of the nonlinear attractor diagram of an optomechanical system where the dominant nonlinearity is that of the radiation pressure interaction.  Due to the limited drive amplitude of the electro-optic modulator used in this work ($\betaEOM \lesssim 3.5$), we are limited to exploring only the lowest red-side attractor.  With the ability to apply larger drives, or to rapidly detune the laser, it should be possible to reach higher-lying islands, and more fully explore the attractor diagram shown in \Figref{waterfall}a.  Further understanding of the latching effects in these measurements should also pave the way to exploiting them for use in metrology experiments as the dynamics that govern whether the oscillator stably latches into an attractor can be a very sensitive function of the oscillator's displacement~\cite{marquardt_dynamical_2006}, thus yielding a precise measurement of the oscillator's environment or state. This latching also allows for systems with memory due to the hysteretic nature of the nonlinearity, as in \cite{bagheri_dynamic_2011,m._i._dykman_theory_1979,siddiqi_rf-driven_2004,siddiqi_direct_2005,jung_multistability_2014}.  Finally, in future devices where the optomechanical coupling rate is larger, these same nonlinearities can lead to quantum mechanical effects which have thus far only been explored theoretically~\cite{ludwig_optomechanical_2008}.

\begin{acknowledgments}
This work was supported by the DARPA ORCHID and MESO programs, the Institute for Quantum Information and Matter, an NSF Physics Frontiers Center with support of the Gordon and Betty Moore Foundation, the AFOSR through the ``Wiring Quantum Networks with Mechanical Transducers'' MURI program, and the Kavli Nanoscience Institute at Caltech.  F.M. acknowledges an ERC Starting Grant OPTOMECH, ITN cQOM.
\end{acknowledgments}


\end{document}